\documentclass[12pt]{article}
\usepackage{amsmath,amssymb}
\usepackage{graphicx,color}
\numberwithin{equation}{section}
\usepackage{cite}
\usepackage{bm}
\usepackage{dcolumn}
\newcommand{\be}{\begin{equation}}
\newcommand{\bea}{\begin{eqnarray}}
\newcommand{\eea}{\end{eqnarray}}
\newcommand{\ba}{\begin{array}}
\newcommand{\ea}{\end{array}}
\newcommand{\ee}{\end{equation}}

\expandafter\ifx\csname mathbbm\endcsname\relax

\else

\fi \textheight 22cm \textwidth 15cm \topmargin 1mm
\begin{document}
\begin{titlepage}
\vspace{10mm}
\begin{flushright}
 IPM/P-2012/040 \\
\end{flushright}

\vspace*{20mm}
\begin{center}
{\Large {\bf Conformally Lifshitz solutions from  Horava-Lifshitz Gravity}\\
}

\vspace*{15mm}
\vspace*{1mm}
{Mohsen Alishahiha$^{a}$ and  Hossein Yavartanoo$^{b}$ }

 \vspace*{1cm}

{\it ${}^a$ School of Physics, Institute for Research in Fundamental Sciences (IPM)\\
P.O. Box 19395-5531, Tehran, Iran \\ }
 \vspace*{0.5cm}
{\it ${}^b$ Department of Physics, Kyung-Hee University, Seoul 130-701, Korea }

\vspace*{0.5cm}

E-mails: alishah@ipm.ir, yavar@khu.ac.kr

\vspace*{2cm}
\end{center}

\begin{abstract}

We show that the IR action of the healthy non-projectable  Ho\v{r}ava-Lifshitz (HL)  gravity
and its small modification exhibit  asymptotically  Lifshitz and  hyperscaling violating solutions,
respectively.
 The model may also have an AdS$_2\times R^d$ vacuum solution. By making use of our results, one may wonder that  within 
the context  of HL gravity the Lifshitz fixed point might be treated as a
CFT deformed by the time component of a vector operator. 
\end{abstract}

\end{titlepage}

\section{Introduction}

Lifshitz fixed point is a critical point where the space and time scale differently.  For spatially 
isotropic scale invariant case, the corresponding fixed point is characterized by a dynamical exponent, 
$z$, as follows
\be
t\rightarrow \lambda^z t,\;\;\;\;\;\;\;\;x_i\rightarrow\lambda x_i,\;\;\;\;\;\;\;i=1,\cdots,d.
\ee
In the context of AdS/CFT correspondence\cite{Mal}, the gravity dual of the Lifshitz fixed point 
is provided by the following metric\cite{KLM}\footnote{See also \cite{Koroteev:2007yp} for an earlier work on a geometry with the Lifshitz scaling.}
\be\label{metric0}
ds^2=-r^{2z}dt^2+\frac{dr^2}{r^2}+r^2\sum_{i=1}^{d}dx_i^2.
\ee
This metric whose isometry group contains the generators of  spatial rotations and  translations, time
translations and dilataions is known as the Lifshitz geometry. Note that the isometry group 
of the Lifshitz geometry does not mix  time and space coordinates.

More generally one may consider a metric which is conformally Lifshitz. The corresponding geometry  may be written as follows\cite{Kiritsis} 
\be
\label{metric1}
ds_{d+2}^2=r^{-2\frac{\theta}{d}} \left(-r^{2z}{dt^2}+\frac{dr^2}{r^2}+r^2{\sum_{i=1}^ddx_i^2}
 \right),
\ee
where the constant $\theta$ is the  hyperscaling violation exponent and the metric is known as 
hyperscaling violating geometry. The reason of this name is as follows.
Since  with a non-zero $\theta$ the distance is not
invariant under the scaling, using AdS/CFT correspondence, 
it indicates  violations of  hyperscaling in the dual field theory. 
More precisely, while in $(d+1)$-dimensional theories 
without  hyperscaling (dual to background\eqref{metric0})  the entropy  
scales as $T^{d/z}$ with temperature, in the present case (dual to background \eqref{metric1}) it scales
as $T^{(d-\theta)/z}$ \cite{{Gouteraux:2011ce},{Huijse:2011ef}}. 

It is obvious that the above metrics are  not  solutions of a pure cosmological Einstein gravity.
This is simply because in the pure Einstein gravity there is nothing to produce an anisotropic in 
the space-time. In fact to obtain such solutions one needs to couple the Einstein gravity  to other
fields. In the minimal case the extra field could be a massive gauge field \cite{Taylor} which can, indeed, produce an anisotropy in the space-time leading to Lifshitz geometry (see also\cite{Tarrio:2011de}). 
More naturally the hyperscaling violating metrics may be found in the Einstein-Maxwell-Dilaton theory
(see for example \cite{{Dong:2012se},{Alishahiha:2012qu}}).

It is important to note that in order to produce the above {\it non-relativistic} solutions,  the models which
have mostly considered in the literature are based on {\it relativistic} actions which are invariant 
under the full space time diffeomorphism. On the other hand taking into account that  conformally Lifshitz geometries  are not invariant under the full Lorentz group, one may naturally pose a question whether  above solutions may be found from 
a non-relativistic action such as that of  Ho\v{r}ava-Lifshitz (HL) gravity\cite{H1,H2}.

Actually  this question has been recently  addressed in\cite{Griffin:2012qx} where the 
authors have shown that the Lifshitz metric can be obtained from HL gravity. Indeed in 
a bottom-up approach using the 
following ADM decomposition of the metric
\be\label{metric}
ds^2=-N^2dt^2+g_{ab}(dx^a-N^adt) (dx^b-N^bdt).
\ee  
the authors of  \cite{Griffin:2012qx} considered  an action as follows
\be\label{action}
S=\frac{1}{2\kappa^2}\int dt dr d^dx\; \sqrt{g} N\left( K_{ab}K^{ab}-\lambda K^2+\beta(R-2\Lambda) +\frac{\alpha^2}{2} \frac{\nabla_a N \nabla^aN}{N^2}   \right),
\ee
where $K_{ab}=\frac{1}{2N}(\partial_t g_{ab}-\nabla_aN_b-\nabla_bN_a)$ is the extrinsic 
curvature of the foliation,
$K=g^{ab}K_{ab}$ and $R$ is the scalar curvature of the metric $g_{ab}$. In the 
context of HL gravity the above  action consists of  the most relevant terms at low energies. Moreover
the gauge symmetries of the action  are the foliation preserving diffeomorphisms which could 
naturally contains the Lifshitz group. 
This is, indeed, the IR action of the healthy non-projectable model introduced in \cite{{Blas:2009yd},{Blas:2009qj}}.

In the present paper we would like to further explore a possibility of having  non-relativistic 
vacuum solutions in  the action \eqref{action}. In particular we shall show that the model exhibits vacua which are asymptotically 
Lifshitz geometry.  Moreover with a small modification, namely assuming an $r$-dependent $\Lambda$, 
the model could support hyperscaling violating solutions. Alternatively the hyperscaling violation
solutions may be obtained by adding a minimally non-relativist scalar field with a wrong sign in 
the potential. Interestingly enough, even though the system is uncharged,  the model also support an AdS$_2\times R^d$ solution which 
might be thought of as an emergent IR fixed point\footnote{Typically an $AdS_2$ factor
appears at the near horizon of the extremal black holes where the backgrounds
are charged.}. This might be expected becuase an  $AdS_2\times R^d$ may be thought of as 
a Lifshitz
geometry with $z\rightarrow \infty$ dynamical exponent.

 Note that for non-zero $\alpha$, AdS$_{d+2}$ is
not a solution of the equations of motion, though setting $\alpha=0$ the equations of motion
admit AdS$_{d+2}$ vacuum solution. It is, however, important to mention that even though  
with $\alpha=0$ we get an  AdS solution, the gravitational model is not the standard GR. This 
is due to the fact that the model does not have the general temporal diffeomorphism.

The paper is organized as follows. In the next section we will obtain asymptotically Lifshitz solution for the action (\ref{action}).
Then assuming an $r$-dependent $\Lambda$ (or by coupling the gravity to a scalar field) we 
find hyperscaling violating solutions. Then we conclude the paper in the section three where we 
shall argue that  within 
the context  of HL gravity the Lifshitz fixed point might be treated as a
CFT deformed by the time component of a vector operator.

\section{Non-relativistic  Solutions From HL Gravity}

\subsection{Lifshitz Geometry}

In this section we would like to study a general solution of HL gravity which is  asymptotically
Lifshitz geometry. To begin with we consider the following ansatz
\be\label{ans}
 ds^2=e^{2f(r)}dt^2+\frac{dr^2}{e^{2h(r)}}+e^{2l(r)}dx_i^2.
 \ee
Of course with $\alpha=0$, even though we have still two more free parameters,  one would not expect 
to get Lifshitz solution from the action \eqref{action}, while
when $\alpha$ is non-zero  this minimal model  has a chance to have such a vacuum. 

 For 
the ansatz \eqref{ans}  one has
\be
N=e^{f(r)},\;\;\;\;g_{rr}=\frac{1}{e^{2h(r)}},\;\;\;\;g_{ii}=e^{2l(r)},\;\;\;\;N_a=0,\;\;\;\;g_{ij}=0,\;\;{\rm for}\;\;
i\neq j.
\ee
Thus
\be
\sqrt{g}=e^{dl(r)-h(r)},\;\;\;\;\;\;K_{ab}=K=0.
\ee
Note also that
\be
R=-de^{2h(r)}\bigg[2l''(r)+(d+1)l'^2(r)+2l'(r)h'(r)\bigg].
\ee
Plugging this ansatz into the  action \eqref{action} and after integration by parts one arrives at
\be\label{act}
S=\frac{{d\beta v}}{2\kappa^2}\int dr \;e^{dl+h+f}\bigg[(d-1)l'^2+2l'f'+\delta f'^2-2\frac{\Lambda}{d} e^{-2h}\bigg]
\equiv\frac{{d\beta v}}{2\kappa^2}\int dr \;e^{dl+h+f} {\cal L}_0
\ee
where $v=\int dt d^dx$ and $\delta=\frac{\alpha^2}{2d\beta}$. To find the functions  $f,l$ and $h$,
one may write down the equations of motion of the above one dimensional action. Furthermore, the equations of motion  are supplemented by a zero energy constraint. 

The corresponding equations of motion of $l$ and $f$ are
\be\label{eom1}
\bigg[e^{dl+h+f}\bigg(\delta f'+l'\bigg)\bigg]'=\frac{1}{2}e^{dl+h+f}{\cal L}_0,\;\;\;
\bigg[e^{dl+h+f}\bigg(f'+(d-1)l'\bigg)\bigg]'=\frac{d}{2}e^{dl+h+f}{\cal L}_0.
\ee
On the other hand since the derivative of $h$ does not appear in the action (\ref{act}), its equation of motion gives us the following constraint  
\be\label{cos1}
{\cal L}_0=-4\frac{\Lambda}{d}e^{-2h},
\ee
which is, indeed, the zero energy constraint.

By making use of the zero energy constraint \eqref{cos1}, the equations of motion \eqref{eom1} may be simplified 
as follows
\be\label{Eq1}
\bigg[e^{dl+h+f}\bigg(\delta f'+l'\bigg)\bigg]'=-2\frac{\Lambda}{d}e^{dl-h+f},\;\;\;
\bigg[e^{dl+h+f}\bigg(f'+(d-1)l'\bigg)\bigg]'=-2 \Lambda e^{dl-h+f},
\ee
One may further simplify the above equations to get
\be\label{Eq2}
(1-d\delta)f'-l' =ce^{-dl-h-f}
\ee
where $c$ is a constant of integration.

Of course, in general, it is not an easy task to solve these 
equations. Nevertheless, inspiring by the known solutions in the literature, one  may guess an 
ansatz with free parameters and try to fix the parameters by plugging it in above equations.
To proceed we will consider the following ansatz
\be
f=z \ln r+\frac{1}{2}\ln\xi(r),\;\;\;\;\;h=\ln r+\frac{1}{2}\ln\xi(r),
\;\;\;\;\;l=\ln r .
\ee
Plugging this ansatz into the equations of motion one observes that there is a non-trivial
solution if we take
\be\label{sol1}
\delta =\frac{z-1}{d z },\;\;\;\;\;\;\;\;\Lambda =-\frac{(d+z-1 ) (d+z )}{2}
\ee
Therefore the solution reads
\be\label{sol2}
ds^2=-r^{2z}\xi(r) dt^2+\frac{dr^2}{r^2\xi(r)}+r^2dx_i^2,\;\;\;\;\;\;\;\;\xi(r)=1-\left(\frac{r_0}{r}\right)^{d+z},
\ee
which is a black brane solution in an asymptotically Lifshitz geometry. 

\subsection{Hyperscaling Violating Metric}

In this subsection we would like to investigate a possibility of having   conformally Lifshitz solution in 
HL gravity. It is clear that an action in the form of \eqref{action} 
cannot exhibit conformally 
Lifshitz solution. This is due to the fact that although the action has an anisotropic nature which is needed
for Lifshitz geometry, it does not have enough free parameters to support the degree of freedom  needed
to have a 
hyperscaling violation solution. Nevertheless one may have a simple modification of the action \eqref{action} 
which could support a hyperscaling violation metric as its vacuum solution.

Actually  one could still work with the same action as \eqref{action} with an assumption that the
cosmological constant $\Lambda$ is not, indeed, a constant. In other words we may assume an $r$-dependent
$\Lambda$. In this case, essentially, the equations of motion are the same as that in the previous
subsection, except one should replace $\Lambda\rightarrow\Lambda(r)$.

With this assumption we will proceed with the following ansatz
\be
f=(z-\frac{\theta}{d})\ln r+\frac{1}{2}\ln\xi(r),\quad h=(1+\frac{\theta}{d})\ln r+\frac{1}{2}\ln\xi(r),\quad l=(1-\frac{\theta}{d})\ln r.
\ee
Plugging this ansatz in the equation \eqref{Eq2} one finds
\be
\xi(r)=1-\left(\frac{r_0}{r}\right)^{d+z-\theta},\;\;\;\;\;\;\;\delta =\frac{z-1}{dz-\theta}.
\ee

On the other hand using the equation \eqref{Eq1} one realizes that the above solution 
is  consistent if 
\be
\Lambda(r)=-\frac{  (d+z-1-\theta) (d+z-\theta) }{2 } \;r^{2\frac{\theta}{d}}.
\ee
This solution represents a black brane in an asymptotically hyperscaling violating geometry\cite{Dong:2012se}. It is clear that for $\theta=0$ the above solution reduces to (\ref{sol2}) in the previous subsection.

Alternatively the hyperscaling violating geometry may be obtained from HL gravity if we 
couple the action \eqref{action} to a non-relativistic matter. In the minimal model the corresponding 
matter field could be a real scalar field whose action consists of 
 a quadratic kinetic term with the right symmetries and  a potential.  In general 
the action for a scalar field may be written as \cite{Kiritsis:2009sh}
\be
S_{Sc}=\frac{1}{2\kappa^2}\int d^d x dt dr \sqrt{g} N\left(\frac{1}{N^2}(\dot{\Phi}-N^a\partial_a\Phi)^2-V(\partial_a \Phi, \Phi)\right)
\ee
Therefore adding this scalar action with the gravity sector \eqref{action}  would provide a model in which we will seeking a  hyperscaling violation solution\footnote{ Note that in this case we can absorb the
cosmological constant in the definition of potential of the scalar field.}. To find such a solution we will consider the following potential for the scalar field
\be
V(\partial_a\Phi,\Phi)=-\frac{1}{2}g^{ab}\partial_a\Phi\partial_b\Phi-V_0e^{\gamma\Phi},
\ee 
note that the first term, comparing with the conventional relativistic scalar field, has ``wrong sign''.  For a static ansatz for the scalar field and ansatz  \eqref{ans}  for the metric, the total action is reduced to the following one dimensional action
\bea\label{act2}
S&=&\frac{{d\beta v}}{2\kappa^2}\int dr \;e^{dl+h+f}\bigg[(d-1)l'^2+2l'f'+\delta f'^2
+\frac{1}{d\beta}\Phi'^2+\frac{V_0}{d\beta} e^{\gamma\Phi-2h}\bigg]\cr 
&\equiv&\frac{{d\beta v}}{2\kappa^2}\int dr \;e^{dl+h+f} {\cal L}.
\eea
Therefore the corresponding equations of motion of $l$ and $f$ are
\be
\bigg[e^{dl+h+f}\bigg(\delta f'+l'\bigg)\bigg]'=\frac{1}{2}e^{dl+h+f}{\cal L},\;\;\;
\bigg[e^{dl+h+f}\bigg(f'+(d-1)l'\bigg)\bigg]'=\frac{d}{2}e^{dl+h+f}{\cal L}.
\ee
On the other hand the equation of motion for $h$  gives us the following constraint 
\be
{\cal L}=2\frac{V_0}{d\beta}e^{\gamma\Phi-2h},
\ee
while for the scalar field  one finds
\be
\bigg[e^{dl+h+f}\Phi'\bigg]'=\frac{V_0\gamma}{2} e^{dl-h+f+\gamma\Phi}.
\ee
Then it is  easy to check that the  ansatz
\be
f=(z-\frac{\theta}{d})\ln r+\frac{1}{2}\ln\xi(r),\quad h=(1+\frac{\theta}{d})\ln r+\frac{1}{2}\ln\xi(r),\quad l=(1-\frac{\theta}{d})\ln r,\quad \Phi=\phi_1 \ln r
\ee
solves the equations of motion if we take
\be
\delta =\frac{z-1}{dz-\theta},
\;\;\;\;\;V_0=(d+z-\theta-1) (d+z-\theta) \beta,\;\;\;\;\;
\gamma^2=\frac{4 \theta}{d(d+z-\theta-1) \beta }
\ee
and then
\be
\xi(r)=1-\left(\frac{r_0}{r}\right)^{d+z-\theta},\;\;\;\;\;\;\;\phi_1=\frac{2\theta}{d\gamma}.
\ee
Therefore the solution is
\be
ds^2=r^{-2\frac{\theta}{d}}\left(-r^{2z}\xi(r)+\frac{dr^2}{r^2\xi(r)}+r^2\sum_{i=1}^ddx_i^2\right),\;\;\;\;\;\;\;
\Phi=\frac{2\theta}{d\gamma}\ln r.
\ee

It is worth to note  that, at least, when the matter field is given by a minimally coupled 
non-relativistic scalar field, the hyperscaling violating solution does exist if the potential 
term of the scalar field has a wrong sign.  Of course this is consistent with the fact that
the hyperscaling violating metric is a solution of pure gravity ( the action \eqref{action})  with the
$r$-dependent $\Lambda$. This means that the matter field should mimic the nature of
a cosmological constant. Then it is  natural to study  the stability of the solution, though we will 
postpone it to a further study.
On the other hand  we had worked with the plus sign in front of $(\partial_a\Phi)^2$ 
term in the potential, the value for $\gamma$ would have been imaginary, leading to 
an imaginary solution.

Moreover the  running of scalar field might suggest that the solution is not IR complete and
therefore one cannot trust the solution for small $r$. In fact the situation is 
similar to the relativistic case studied in \cite{Bhattacharya:2012zu} where it was shown that
by taking into account the quantum corrections the  
magnetically charged hyperscaling violating solution could be completed at IR by an
emergent AdS$_2\times R^d$ solution. 

Although the aim of the present paper is not to explore a possible IR completion of the geometry, it is
important to note that with a non-zero $\alpha$ the model admits another interesting solution 
as follows
\be
\Phi=\Phi_0,\;\;\;\;\;\;\delta=\frac{1}{d},\;\;\;\;\;\;V_0=-1,\;\;\;\;\;\;\;\gamma=0.
\ee
Therefore the corresponding  metric is
\be
ds^2=-r^2dt^2+\frac{dr^2}{r^2}+\sum_{i=1}^ddx_i^2,
\ee
which is an  AdS$_2\times R^d$  solution.  It is very interesting that HL gravity
with non-zero $\alpha$ admits an AdS$_2\times R^d$  solution, though it may have AdS$_{d+2}$ 
only if $\alpha=0$. Of course it can be understood from the fact that the AdS vacuum is 
an isotropic solution.

Having found an AdS$_2\times R^d$  solution, one may wonder that this solution can be 
thought of as an emergent IR fixed point which could complete the geometry even though we are
dealing with the HL gravity.

\section{Conclusions}
 
In the previous section we have shown  that the minimal model defined by the 
action \eqref{action} admits asymptotically Lifshitz vacua. 
 An interesting observation is that the resultant solution (the equations \eqref{sol2}
and \eqref{sol1}) does not depend on 
the parameter $\lambda$. Moreover the parameter $\beta$ can be absorbed by a 
redefinition of $\alpha$. Though the solution does crucially depend on $\alpha$.

Indeed, as long as $\alpha$ is non-zero, the metric \eqref{sol2} is a 
solution of the equations of motion for arbitrary $\lambda$ and $\beta$. In particular 
one may set $\lambda=\beta=1$, so that  the first terms of the action \eqref{action} reduce to
the standard Einstein-Hilbert action. More precisely setting 
\be
G_{\mu\nu}=\begin{pmatrix}
         -N^2+N^aN_a & N_a\\
        N_b & g_{ab}\end{pmatrix},
\ee
the action \eqref{action} may be recast into the following form
\be\label{model}
S=\frac{1}{2\kappa^2}\int dt dr d^dx\; \sqrt{-G}\left( {\cal R}-2\Lambda +\frac{\alpha^2}{2} \frac{\nabla_a N \nabla^aN}{N^2}   \right),
\ee
where ${\cal R}$ is the scalar curvature of the metric $G_{\mu\nu}$.

It is  important to note that when we set $\alpha=0$ although
the action gets the same form as the Einstein-Hilbert one, due to the missing of 
temporal diffeomorphism, the model does
not reduce to standard GR theory. Nevertheless as it is evident from the equations \eqref{sol1}
and \eqref{sol2}  when $\alpha$ is zero the natural vacuum of the model is an  AdS geometry.
While for non-zero $\alpha$ the natural 
solution is the one with an anisotropic direction such as the Lifshitz geometry. 

If we think of the model given by the action \eqref{model} as a gravity which provide a
gravitational description for a non-relativistic field theory,  then having the $\alpha$ term 
might be thought of as perturbing the dual theory by an operator which induces an RG flow
from an isotropic CFT vacuum into a vacuum with an anisotropic along the time direction.

To explore this point better, following \cite{Blas:2009yd}  we utilize the St\"uckelberg formalism
by which the general covariance may be restored by introducing a scalar field $\varphi$, known
as the khronon. In this
formalism the surfaces of foliation are defined by  $\varphi={\rm constant}$. Then it is useful
to define a unit normal vector, $u_\mu$,
\be
u_{\mu}=\frac{\partial_\mu\varphi}{\sqrt{G^{\mu\nu}\partial_\mu\varphi\partial_\nu\varphi}},
\ee
by which the action \eqref{model}  may be generalized to
\be\label{cov}
S=\frac{1}{2\kappa^2}\int dt dr d^dx\; \sqrt{-G}\left( {\cal R}-2\Lambda +\frac{\alpha^2}{2} 
a_\mu a^\mu  \right),
\ee
where $a_\mu=u^\nu\nabla_\nu u_\mu$. 

 Then it is  possible to compensate the 
non-invariance of HL  gravity under temporal diffeomorphisms by a suitable transformation of 
$\varphi$. Note that the HL gravity can be obtained by  a foliation defined by $\varphi=t$ which 
is known ``unitary gauge''. In other words, the action may be thought of as 
a deviation  of Einstein gravity\footnote{It is, however, important to note that setting $\alpha=0$, will remove the kinetic term of the scalar field from the action  and therefore the limit is degenerate
\cite{Blas:2009yd}. Nevertheless we will consider the $\alpha$ term as
 a sufficiently small perturbation.}. Deforming the action by the $\alpha$ term  in the gravity side
 would correspond to perturbing the dual CFT.

In general it might be difficult to make the above statement precise. Namely it is not clear how to identify
the corresponding dual operator.  Heuristically, one may treat $a_\mu$ as a vector field in the bulk
by which the action is deformed. Then there would be a vector operator in the dual theory which 
perturbs the corresponding CFT. In particular if we consider  the time component of the dual operator
to be non-zero then in the bulk gravity we have to deform the theory by $a_t$ which in turn is
equivalent to $\varphi=t$ foliation. In this case the end point of the RG flow will be a Lifshitz fixed point.
It is an easy exercise to see this procedure in the covariant form.  Indeed from the ansatz \eqref{ans} one has
\be
{\cal R}=-2e^{2h}\left[d l''+f''+\frac{d(d+1)}{2}l'^2+f'^2+d l'f'+d l'h'+f'h'\right],\;\;
\sqrt{-G}=e^{dl-h+f}.
\ee
Plugging these expressions in the above action and performing an integration by parts and setting
$\varphi=t$ one  arrives at the same one dimensional action of that in the previous section whose natural vacuum 
is the asymptotically Lifshitz metric.

More generally  we may perturb the dual CFT by a spatial  component of the dual operator, let say 
along $x_1$ direction;  ${\cal O}_1$. Then the gravity will be deform by $a_{x_1}$. 
A natural way to get the desired field is to foliate the space time by $\varphi=x_1$. More 
explicitly one has
\be
ds^2=-A^2 dx_1^2+g_{mn} (dx^m-B^m dt)(dx^n-B^ndt),\;\;\;\;\;\;\;m,n=t,2,\cdots, d.
\ee
Using a proper ansatz and plugging it into the covariant form of the action \eqref{cov}, one 
can  show that asymptotically the resultant metric has 
the following  anisotropic scaling
\be
r\rightarrow \lambda^{-1}r,\;\;\;\;\;x_1\rightarrow \lambda^z x_1,\;\;\;\;(t,x_i)
\rightarrow \lambda (t,x_i),\;\;{\rm for}\;i=2,\cdots,d,
\ee
with  $z=2/(2-\alpha^2)$. More generally one finds 
\be
ds^2=-r^2 dt^2+r^2 \zeta dx_1^2+\frac{dr^2}{r^2\zeta}+r^2\sum_{i=2}^d dx_i^2,\;\;\;\;\;\;
\zeta=1-\left(\frac{r_0}{r}\right)^{d+z}.
\ee

Therefore one may conclude that  holographically the Lifshitz fixed point may be considered 
as an AdS solution perturbed by  the time component of the vector operator ${\cal O}_\mu$
\footnote{In the context of AdS/CFT correspondence a similar idea-rather more precise- has been
explored by K. Skenderis et al.\cite{TO}}.
Other component of the operator perturbs the theory to another vacuum with an anisotropic 
in a spatial direction.

Another interesting observation we made in this paper is that the  HL gravity we have been considering 
exhibits an IR emergent fixed point where the background develops an AdS$_2$ geometry. In other 
words we find a vacuum with AdS$_2\times R^d$ geometry. It is interesting in the sense that we arrive
at an $AdS_2$ geometry even though the background metric is not charged.

We have also shown that with a small modification, namely assuming an $r$-dependent 
$\Lambda$ the theory could also support vacua with hyperscaling violating geometry. 

Finally we note  that it should  be natural to couple the model to a non-relativistic gauge field and find the
charged conformally Lifshitz solution.

\section*{Acknowledgments}

We would like to thank D. Allahbakhshi, A. Mollabashi, M. Mohammadi Mozaffar, M. R. Tanhayi and
A. Vahedi  for useful discussions. The work of H.Y  is supported by the National Research Foundation of Korea Grant funded by the Korean Government (NRF-2011- 0023230).

\end{document}